\documentstyle[aps]{revtex}
\def\gtlt{{\raise2.5pt\hbox{$>$}\kern-8pt\lower2pt\hbox{$<$}}}
\tighten
\begin{document}
\draft
\preprint{UCF-CM-94-001}
\title
{Nonlinear Steady-State Mesoscopic Transport, I. Formalism}
\author{M.D. Johnson and O. Heinonen}
\address{
Department of Physics, University of Central Florida, Orlando, FL 32816-2385
}
\date{March 4, 1994}
\maketitle
\begin{abstract}
We present an approach to steady-state mesoscopic transport based on the
maximum entropy principle formulation of nonequilibrium statistical mechanics.
This approach is valid in the nonlinear regime of high current,
and yields the quantization observed in the integer quantum
Hall effect at large currents.  A key ingredient of this approach,
and its success in explaining high-precision Hall measurements,
is that the occupancy of single-electron states depends
on their current as well as their energy.  This suggests
that the reservoir picture commonly used in
mesoscopic transport is unsatisfactory outside of
linear response.
\end{abstract}
\bigskip
We have recently developed an approach to steady-state mesoscopic transport
not restricted to the linear response regime.  By nonlinear here we mean that
the driving force (voltage or current) is large---too large for linear
response calculations, but well within the range of typical experimental
values.
Our approach is applicable to quasi-one-dimensional systems and to
two-dimensional systems in strong magnetic
fields, such as those exhibiting the integer quantum Hall effect.
This work was reported in a brief form elsewhere\cite{HJ1}.
Our study was motivated by the important but often neglected
fact that the integer quantum Hall effect (IQHE) \cite{IQHE} is exhibited
even when the currents and voltages are very large \cite{NIST}.
Because the magnetic field strongly suppresses inelastic scattering,
systems exhibiting the IQHE can be viewed as mesoscopic even though
they are relatively large \cite{suppression}, and the IQHE itself can
be viewed as a near-ideal manifestation of mesoscopic transport.
The IQHE at low currents is well understood in terms of the Landauer-B\"uttiker
(LB) approach \cite{Landauer,Buttiker} to mesoscopic transport.
This approach, however, appears to be fundamentally a linear response
theory \cite{ADStone,BarangerStone}.  (See, however,
\onlinecite{VanSon,LandauerGW}.)
When applied nonetheless at high currents, it fails to yield the observed
quantization.
The quantization at high currents appears to us to be an extraordinary
phenomenon.  It is easy to imagine many things that can end quantization
(dissipation, backscattering, etc.), but it is not obvious how to
restore quantization.
It is perhaps possible that quantization at
high current might result from conventional approaches to nonequilibrium
transport (such as the Keldysh or quantum Boltzman approaches \cite{Ferry});
but these are difficult even near linear response and their behaviors
at large currents simply unknown.  The high-current quantization is
so extraordinary that it seemed likely to us that a successful theory
of large-current mesoscopic transport would have to take its highly
non-equilibrium nature into account from the very beginning.

We found an apparently fruitful direction in the
maximum entropy approach (MEA) to nonequilibrium statistical mechanics
\cite{MEP},
in which the density matrix is found by maximizing the information entropy
of the system, subject to constraints which fix the expectation values of
observables.   This approach should in principle be quite generally
applicable to equilibrium and nonequilibrium statistical mechanics,
but in fact there have been very few examples of the latter.  For reasons
which we shall explain below, mesoscopic systems (including those
exhibiting the IQHE) are ideally suited for the MEA, and our calculations
are exact for mesoscopic systems consisting of non-interacting electrons.
This gives as a first benefit an exact Hall quantization at zero
temperature, for ideal systems, for almost arbitrarily large currents.
A consequence of this work is that it suggests that the picture
of current and voltage probes as {\it reservoirs}---which underlies nearly all
approaches to transport in
mesoscopic systems, and which has proven extremely successful at
low currents---may not be satisfactory at high currents.

In the present and a companion paper
we present our theory in detail together with discussions and
applications. The present paper, part 1, develops the formal theory,
and the companion, part 2, contains various applications.  In the
present paper, section I contains a more detailed
introduction and motivation.  Section II contains our maximum
entropy approach together with a detailed example, and section III contains
a discussion of the maximum entropy approach, the resulting electronic
distributions, and the relationship between this work and
more conventional approaches.

\section{Introduction; Landauer-B\"uttiker formalism}

A mesoscopic system typically consists of a `device' (such as a Hall bar or
quantum wire) and a number of current and voltage probes.  The device itself is
truly mesoscopic, by which we mean that the length of the device
is smaller than the phase-breaking length. The connection between the
device and the external world is provided by the
probes, which are attached to the device at terminals.
(In reality, the
terminals consist of, for example, the $n^+$-doped regions which connect
to the inversion layer in a silicon metal-insulator field-effect transistor,
or the `fingers' of indium diffused through the different layers
in a heterostucture.
Thus, in a real device carriers pass into the device through
terminals which are no larger than the device itself. The carriers in general
will suffer both elastic and inelastic scattering, and so dissipate energy, in
the terminals.)
We presume that within the mesoscopic device itself scattering is
entirely elastic and dissipationless.

It is not obvious how to handle the complicated system of device plus probes.
The greatest advance in understanding mesoscopic transport came from an
approach
originally due to Landauer \cite{Landauer,LandauerGW,Imry,Buttiker}, in which
the conduction of the entire sample is treated as a scattering problem.
There are two central concepts in this model.
The first is the `reservoir' --- the probes are treated as macroscopically
large (essentially infinite) reservoirs which inject carriers into the device
through ideal leads.  It is assumed that
each reservoir is in local equilibrium described by a local
chemical potential $\mu_m$, and that
the occupancy of electronic states entering the device from a reservoir
is determined by the local chemical potential of that reservoir.
We will use the term `reservoir' only in this restricted sense.
More generally we will refer to a source of carriers as a `terminal'.
The second key concept is that the motion of carriers through the device itself
is treated as an elastic scattering problem.  Carriers entering from a
reservoir
into the device are scattered either back into
the original reservoir or outward into the other reservoirs.  The scattered
electrons then equilibrate deep within the reservoir with the electrons
in the reservoir.  Scattering in the terminals randomizes the energy
and phase of the carriers, which eliminates any quantum
interference \cite{ButtikerRev}.

To formulate the scattering problem one needs asymptotic regions in the
leads in which states carry electrons either away from or towards the
device \cite{ButtikerRev} (and in which evanescent modes have decayed away).
Such leads are quasi-one-dimensional, and states in them can be labeled
by subband index $n$ and wavenumber $k$.  (In the presence of a magnetic
field, $n$ is a Landau level index.)  Within the system consisting of
the device plus the asymptotic regions scattering is elastic.
This gives the conventional treatment of mesoscopic devices:
conduction occurs as elastic scattering of carriers injected into the
asymptotic regions from reservoirs \cite{ButtikerRev}.
In this paper we will adhere entirely to the scattering viewpoint.
We will argue, however, that outside the linear regime one cannot
treat the terminals (which inject carriers) as reservoirs in the
specific sense defined above.

To be definite, we will use the following notation and conventions.
For simplicity, we
assume that the electrons are noninteracting and spinless.  (Both
restrictions can be removed---spin just adds another label, and the formulation
we give later can be extended to the case of interacting many-body states.)
In an $M$-terminal device we
denote the probes by $m=s,d,2,\ldots,M-2$.  Here $s$ denotes the source of
current
and $d$ the drain, since in typical experiments current flows in one lead
and out one other.
In general we denote a complete orthogonal set of single-particle eigenstates
by $|\psi_\alpha\rangle$.  These have energies $\epsilon_\alpha$ and
carry net currents $i_{m,\alpha}$ through terminal $m$.  (A positive
$i_{m,\alpha}$
corresponds to positive current flow {\it into} the device from terminal $m$.)
The scattering picture mentioned above can be made precise by supposing
that the leads can be treated as semi-infinite and straight \cite{ButtikerRev}.
In this case, a particularly useful set
of eigenstates for multi-terminal ($M>2$) systems
are the scattering states \cite{BarangerStone,Sols}
$|\psi^+_{mnk}\rangle$ ({\it i.e.}, $\alpha=mnk$).
The state $|\psi^+_{mnk}\rangle$ is
incoming into the device from terminal $m$; $n$ and $k$ denote
the asymptotic wavenumber and subband index of the incoming wave.
The state has energy $\epsilon_{mnk}$ and
carries current $i^0_{mnk}$ into the device at terminal $m$.
(Landauer objects to this lead geometry as incompatible with
the reservoir concept \cite{LandauerGW}.  But our discussion of
the LB approach needs only a subband index and wavenumber, and
does not use this geometry at all.  We have specified this
geometry here because the scattering language it makes precise
is useful for the subsequent sections.)
In general, after scattering within the device, the state carries
outward current through each terminal.  We denote by $i_{m',mnk}$
the {\it net} current of this state
into the device at terminal $m'$, with the convention that current flowing {\em
into}
the device is positive.
The state's net current at $m'$
is related to its incoming current at $m$ by
$i_{m',mnk}=i^0_{mnk}(\delta_{m'm}-\sum_{n'k'} t_{m'n'k',mnk})$,
where $t_{m'n'k',mnk}$ is the transition probability
obtained from the scattering matrix in the $|\psi^+_{mnk}\rangle$
representation (with a proper normalization\cite{Sols}).
The elastic scattering within terminals
which can exist in a real system can be included as contributions
to the transmission probabilities.

The reservoir and scattering concepts underlie the Landauer-B\"uttiker (LB)
theory of mesoscopic transport \cite{Landauer,Buttiker,ButtikerRev}.
A key point in this theory is that voltage as well as current contacts
are treated identically and described as reservoirs \cite{Buttiker}.
The reservoirs enter this theory in several important ways:
they determine the electronic distribution,
they randomize the phase of occupied states (which eliminates interference),
and they provide a prescription for determining
voltage differences.  Combined, these permit one to calculate current-voltage
($I$--$V$)
curves.  First consider the distribution of electrons in the device.
Suppose the $m^{\rm th}$ terminal is a reservoir described by
a local chemical potential $\mu_m$. Then
the states in the attached lead carrying current toward the device are
occupied according to the Fermi-like distribution
$f_{mnk}^{LB} = 1/[e^{\beta(\epsilon_{mnk}-\mu_{m})}+1]$.
(Here $\beta=1/k_BT$, where $T$ is the temperature and $k_B$
is Boltzman's constant.) The net current flowing into the system at, say,
the source is $I=\sum_{mnk}i_{s,mnk}f_{mnk}^{LB}$.  Suppose that in a
multiprobe device the current
flowing in at the source and out at the drain is $I$, with zero net
current at other terminals.  In the LB approach this is enough
information to determine the local chemical potentials $\mu_m$ (within
an additive constant).  Within the reservoir picture, it is then obvious
that the measured voltage difference between two terminals (1 and 2, say)
must be $V=(\mu_1-\mu_2)/e$, where $e$ is the charge of an electron.

Consider the particular case of a two-terminal device at zero temperature,
with some current $I$ flowing from source to drain.
The current is in this picture caused by a voltage difference
$V=(\mu_s-\mu_d)/e=IR$ between source and drain.  Following the standard
LB approach, at low voltages this gives a resistance $R=h/(je^2\widetilde
t\,)$,
where $j$ is the number of occupied subbands
and $\widetilde t$ is the total transmission probability at $\mu_s$.
This resistance is quantized in the absence of backscattering
($\widetilde t=1$).  In real quantum Hall systems, there is a macroscopic
separation between left- and right-moving states, so that in fact
backscattering
is highly suppressed, and indeed $\widetilde t$ is very nearly unity
\cite{backscatter}.
(We have here neglected dissipation that occurs due to contact
resistance at the source and drain.  In a real two-terminal
device this dissipation prevents perfect quantization of resistance.
Resistances can be quantized only in a multi-terminal device when
measured between two terminals through which no net current flows.)
Here we have briefly given the two-terminal version of this explanation,
but following B\"uttiker \cite{Buttiker} it can be generalized to the
multi-terminal case (see below), in which case the
corresponding voltage is the transverse or Hall voltage.
Hence the LB theory gives a very satisfying microscopic explanation of
how the conductance can be so accurately quantized in the
IQHE \cite{ButtikerRev}---at
least in the low-current regime, as we will explain below.

We have summarized the LB theory here in a way appropriate for linear
response---the difference $\mu_s-\mu_d$ is presumed to be small.
This is assumed in nearly all uses of the LB
approach \cite{Buttiker,ButtikerRev,ButtikerAdv}.  For example, we treated
the transmission $\widetilde t$ as a constant, which requires in part that
$\mu_s-\mu_d$ be small.  It is possible to
generalize this to the case where the transmission depends significantly on
energy near the Fermi level \cite{Sivan}, as would be needed if $\mu_s-\mu_d$
is
large.  In the remainder of this section, however, we
will be concerned only with ideal systems (perfect transmission), so this
type of generalization is not relevant here.

The LB theory of macroscopic transport has been used to interpret a wide
variety of experiments (see, for example, a partial listing in
Ref.~\onlinecite{ButtikerAdv}).
In fact, the fundamental model of reservoirs at (local) equilibrium with
local chemical potentials $\mu_m$ has been used in essentially all mesoscopic
calculations, including microscopic methods such as non-equilibrium
Green's functions.  There is no reason to doubt the fundamental soundness
of treating the terminals as reservoirs in the low-current regime.

Despite the many successes of the LB approach, there are important
experiments which it seems unable to explain.  Chief among these are
the actual high-precision IQHE experiments.  In our above application
of the LB approach to the IQHE we did not mention an important point:
the LB theory only predicts quantization in the IQHE when the current
is small.  When the current is large (as large as those typical in
high-precision experiments), the same argument predicts a failure of
quantization.    There is at least no straightforward way to extend the
LB approach to give quantization at large currents.  The culprit
appears to be the very reservoir concept itself.

Let us examine this in detail, first in a ideal two-terminal example.
Denote the minimum energy of the $j^{\rm th}$ subband by $E_j$.
Suppose to begin with that the local chemical potentials of both the source
and drain exceed $E_j$ but are less than $E_{j+1}$.  (See Fig.~\ref{LBfig}(a).)
In an ideal system at zero temperature, the net current in each subband
is, according to the LB approach,  $e(\mu_s-\mu_d)/h$.
(This simple form occurs because of a cancellation
between the density of states and the current carried in each single particle
state. \cite{Halperin}) With $j$ occupied bands, this gives a current
$I=je(\mu_s-\mu_d)/h$.  The two-terminal voltage in the reservoir
picture is $V=(\mu_s-\mu_d)/e$,
and so the two-terminal resistance is $R=V/I=h/(je^2)$.

Now let us suppose that the local chemical potential of the source
(but not of the drain) is increased above $E_{j+1}$, so that the
source injects electrons into the $(j+1)^{\rm st}$ subband, while
the drain does not.  This is shown in Fig.~\ref{LBfig}(b).
In an ideal system at zero temperature, the
net current contributed by each of the lower $j$ subbands is the
same as above, but the net current in the $(j+1)^{\rm st}$ subband
is $e(\mu_s-E_{j+1})/h$.  Thus the total current is
$I=je(\mu_s-\mu_d)/h + e(\mu_s-E_j)/h$.  The voltage difference must still
be $V=(\mu_s-\mu_d)/e$ in the reservoir picture, so in this case the LB
approach
gives a two-terminal
resistance $R=V/I$ which lies between $h/(je^2)$ and $h/[(j+1)e^2]$.
Notice that this happens whenever one or more of the subband minima lie
between $\mu_d$ and $\mu_s$, which must occur whenever $eV$ exceeds the subband
spacing.
That is, even in the case of an ideal system, with no backscattering,
the LB prediction is that the two-terminal resistance should not be quantized
when $eV$ is large.  And yet in the high-precision IQHE experiments
quantization is found to be extremely accurate even when $eV$ is
10 or 100 times the subband spacing \cite{NIST}.
(The same argument has been invoked to explain the large-voltage
failure of resistance quantization in quantum point contact experiments
within the LB formalism \cite{van Houten}.)

The above argument also works in an ideal multi-terminal IQHE system.
Suppose there are four terminals (taken to be identical for simplicity),
with two current terminals $s$ and $d$, and two transverse voltage terminals
$1$ and $2$ (see Fig.~\ref{devicefig}).  Then to get zero net current in
terminal 1, it is necessary that $\mu_1=\mu_s$; similarly, $\mu_2=\mu_d$.
A given $\mu_s$
and $\mu_d$ give the same total source-to-drain current as above, and the
transverse (Hall) voltage is $(\mu_1-\mu_2)/e$, also as above.  Hence
the ideal four-terminal LB results are identical to the ideal two-terminal
measurements: the resistances are quantized at low but not at high currents.

The discussion thus far has treated the electrons as noninteracting.  One might
suppose that a more accurate treatment, still within the LB picture but
incorporating electron-electron interactions, might modify the large-current
results.  At the lowest (Hartree) level of approximation, electron-electron
interactions can cause the subbands to deviate, perhaps significantly, from
the wavenumber dependence which would arise in the noninteracting
case \cite{Halperin}.  This consequence of electrostatics is pictured
schematically in Fig.~\ref{bent}(a), which shows the bending of energy
levels (not too near a contact) in an IQHE sample due to the combined
presence of a Hall field and an edge confinement.
Of itself, this bending of bands has no effect on the argument above.
If states in one level are occupied out to $\mu_s$ and $\mu_d$, then this
level contributes $e(\mu_s-\mu_d)/h$ to the current, regardless of the band's
shape
(because of the cancellation between density of states and current per state).
So as long as all subbands shift in
energy more or less together, as is the case with Hartree interactions,
the picture is unchanged:  a large
voltage causes partial occupancy of higher subbands and hence failure of
quantization.

In the above paragraph we assumed that, despite electrostatics, ideal
reservoirs are able to occupy all states up to energies $\mu_s$ and $\mu_d$
(at zero temperature).  Van Son and Klapwijk have recently made a serious
attempt to examine more closely the consequences of electrostatics for
the LB approach to the IQHE \cite{VanSon}. Their starting point is that
electron states (except very close to the current probes)
can be described by bulk Landau levels (subbands) with a shape given by
a self-consistently determined electrostatic potential (as in Fig~\ref{bent}.)
Central to their analysis is an argument that the source injects electrons
only in a range of energy $\mu_s-\Delta<\epsilon<\mu_s$.
In other words, their source, because of an electrostatic barrier,
does not act as a reservoir for carriers---it
does not occupy all current-carrying states with energies $\epsilon<\mu_s$.
Within some small length from the source, determined by inelastic scattering
near the source, these electrons `relax' to fill (bulk) subbands
up to some energy $\mu_s'$ (with $\mu_s'<\mu_s$).
Consider the case where $\sigma_H=e^2/h$. In this
case, according to Van Son and Klapwijk, electrons injected
at the source relax to fill only states in the lowest (bulk) subband,
regardless of how large $\mu_s$ is. The result is that the lowest subband
is filled from $\mu_d$ (for left-movers) to $\mu_s'<\mu_s$ (for right-movers).
This is pictured in Fig.~\ref{bent}(b).  If this relaxation occurs before
the first voltage probe, they argue, a multi-terminal measurement would
show the ordinary IQHE.  Note that if $\mu_s'$ lies above the minimum of
the next subband, this `relaxation' results in some occupied states
({\it e.g.}, in the lowest subband) lying at higher energies than
empty states in the next subband (as in Fig.~\ref{bent}(b)).

We believe that Van Son and Klapwijk were correct in their conclusion
that the conventional picture of terminals as reservoirs must be
modified somehow at higher currents and voltages.  But there appear to
be some problems with the details of their arguments.
Following B\"uttiker \cite{Buttiker}, Van Son and Klapwijk suppose that the
voltage probes can be treated as reservoirs with zero net current.
Let us consider what happens when $\mu_s'$ is greater than the energy
minimum of the next higher subband, as pictured in Fig.~\ref{bent}(b).  This
certainly would occur if $eV_H\gg\hbar\omega_c$.
If the voltage terminals are  treated as reservoirs, then the first probe
`downstream' from the source is at some electrochemical potential $\mu_1$
(in the geometry of Fig.~\ref{devicefig}).
For simplicity let us assume that nothing complicated happens at the drain so
that (in this geometry) $\mu_2=\mu_d$.
The occupied states after relaxation carry current $I=e(\mu_s'-\mu_d)/h$, and
the Hall voltage is $(\mu_1-\mu_2)/e$.
This gives a quantized resistance only if $\mu_1=\mu_s'$.  But there is no
reason for this to be the case.  Suppose that the voltage terminal acts as
an ordinary reservoir and injects carriers into all states (in {\it all}
subbands) with energy $\epsilon<\mu_1$.   States in the higher subband
would be occupied, and so, to get zero net current at the probe, it would
have to be the case that $\mu_1<\mu_s'$.
Suppose instead that the voltage terminal, like the current source, injects
carriers into only a range of energies, after which the carriers relax.
If the voltage and current probes are identical, then by symmetry zero net
current in the former would require $\mu_1=\mu_s>\mu_s'$.  In neither of
these two natural possibilities would the Hall resistance be quantized.

Second, Van Son and Klapwijk simply assume that the relaxation process
is so efficient that at low temperatures states in the lowest subband
end up filled continuously from $\mu_d$ to $\mu_s'$ (even
though the current source and drain do not directly feed all of them),
while the upper subbands all are empty.  They assume
this to be the case even if $\mu_s$ is high enough to inject carriers
into a higher subband, as pictured in Fig.~\ref{bent}(b).
This might work for voltages slightly above the subband spacing, but it
is harder to believe that there should be no carriers in any of the higher
subbands if $\mu_s-\mu_d$ is 10 or 100 times $\hbar\omega_c$.
Moreover, the relaxation into the lowest subband must be complete (to get
quantization) only as long as $\mu_d$ lies below the minimum of the second
band, as pictured in Fig.~\ref{bent}(b).  Suppose that $\mu_d$ increases
slightly so that it moves above the minimum of the second subband (while
$\mu_s$ is held fixed).  These appear to be the conditions under which
one should expect a conductance of $2e^2/h$: both the
source and drain attempt to occupy states in the next subband.  Then according
to the picture of Van Son and Klapwijk, it must be the case that states in the
second subband also get filled from $\mu_d$ around to some $\mu_s'$, after
relaxation, giving a conductance of $2e^2/h$.  Yet why should relaxation bring
all carriers to the lowest Landau level when $\mu_d$ lies slightly below the
second subband's minimum, and the same relaxation allow carriers to fill two
subband when $\mu_d$ increases to lie
slightly above the minimum of the second subband?
It is difficult to see how this supposition of large-scale
relaxation into lower subbands can be entirely correct in detail.

Let us turn back to the usual reservoir picture, and examine whether
including the interactions more accurately---by, say, including
exchange---can restore the high-current quantization.
The exchange interaction effectively lowers the energy
of the occupied single-particle states relative to unoccupied states, and in
principle this could remove the partial occupancy of higher subbands (the
cause of the failure to quantize).  Suppose for example that at a low
two-terminal voltage $V$ states in only the lowest subband are occupied.
Let the voltage $V$ increase.  Suppose further
that the exchange and correlation energies effectively
lower the energies of the occupied states in this subband so that they all
lie below the minima of higher subbands.  Then the higher bands would
remain unoccupied, and the resistance remain quantized.
The difficulty with this argument is that the quantization in high-precision
IQHE measurements at voltages 10 or 100 times the unperturbed Landau level
spacing \cite{NIST} would require the exchange and correlation energies to
exceed $10\hbar\omega_c$ or even $100\hbar\omega_c$.
In fact the exchange and correlation energies are much smaller (of order
$10^{-3}\hbar\omega_c$ in the fractional quantum Hall effect), and it
seems implausible that this can be used to modify the LB approach to give
the correct quantum Hall resistance at high currents and voltages.

Perhaps the electron distributions change at large current
as a result of some complicated behavior (beyond electrostatic
considerations \cite{VanSon}) at the terminals where the
leads and mesoscopic system are joined.
There is as yet no detailed explanation
of how this could occur.  Such an argument would also be outside the spirit
of Landauer's original idea.  His insight was that it should be possible
to ignore the (perhaps very complicated) details of the terminals, which
are described entirely by the transmission probabilities, and concentrate
on general principles (see for example the discussion
in \onlinecite{LandauerGW}).
The existence of high-current quantization
in the IQHE and other mesoscopic devices argues, we believe, that this
insight is fundamentally correct.
However, based on the experimental evidence of the high-precision measurements
of the quantum Hall resistance, we believe that the model of terminals as
macroscopic reservoirs in local equilibrium is unsatisfactory outside the
regime of linear response.
Here we will present an approach to mesoscopic transport which can be viewed
as being within the spirit of Landauer's idea, in that we assume
it is possible to neglect the details of the terminals.  Our approach
leads, however, to a different occupancy of electronic states (and hence
requires modification of the reservoir concept), and appears capable of
describing transport in mesoscopic devices outside of the low-current realm.

\section{Maximum entropy approach to mesoscopic transport}

The nonlinear steady-state mesoscopic transport theory which we have
developed is based on the maximum entropy approach to nonequilibrium
statistical mechanics.  In this section we will develop this approach to
transport and provide a detailed example which illustrates how it can
explain the IQHE even at large currents and voltages.

\subsection{Density matrix}

The central ingredient of the maximum entropy approach (MEA) is
the information entropy $S_I$ \cite{MEP}.  If a complete set of eigenstates
of a thermodynamic system is labeled by $\gamma$, then the information
entropy is given by
\begin{equation}
S_I=-c\sum_\gamma p_{\gamma}\ln p_{\gamma}
\label{entropy}
\end{equation}
where $p_{\gamma}$ is the probability that the system is in a given
microstate.  Here $c$ is an unspecified constant.  (When the MEA is
applied to equilibrium thermodynamics, it can be shown that the
information and thermodynamic entropies are identical when $c$
is chosen to be Boltzman's constant $k_B$.  We will see that this
is also true in the case of steady-state mesoscopic transport.)
As in any
thermodynamic calculation, the first necessity is to define `the system'.
We assume that we can define the system to be the device (including
the asymptotic leads),
as described earlier. In general, $\gamma$ then simply
refers to a many-body electron state. In the case of noninteracting
electrons each such microstate corresponds to a particular set
of occupied single-particle scattering states. It is the fact that the device
itself is mesoscopic which allows for
a straightforward description of the microstates. In the presence of
dissipation, this is not so easy. In fact, in spite of claims to
its general applicability, nearly all applications of the MEA to
dissipative nonequilibrium systems have been limited to expansions about
equilibrium \cite{MEP}.

The fundamental postulate of the MEA is that the probabilities
$p_{\gamma}$ are those which maximize the information
entropy---subject to constraints which describe certain given
or known observables.  The method itself does not give a
prescription for determining what are the constraints.
These must be determined from physical considerations.
(We will discuss this point in more detail later.)
In the case of equilibrium thermodynamics, it is assumed that the
internal energy $U$ and electron number $N$ can be taken as given,
whether or not they are actually measured.  In the case of
steady-state transport one knows in addition (by measurement)
the net current at each terminal.  We therefore include this
as an additional constraint and maximize
the information entropy subject to the constraints
$\langle \hat H \rangle = U$, $\langle \hat N \rangle = N$,
and $\langle {\hat I}_m \rangle = I_m$.  Here
$\hat H$ is the Hamiltonian, $\hat N$ is the
particle number operator, and ${\hat I}_m$ is operator
giving the net current in lead $m$ \cite{currentoperator}.
These constraints are conveniently imposed using Lagrangian
multipliers.
The probability entering Eq.~(\ref{entropy}) can be written as the matrix
element $p_{\gamma}=\langle \gamma | \hat\rho | \gamma \rangle$
of the density matrix $\hat\rho$.  Then maximizing $S_I$ subject
to the constraints gives the density matrix
\begin{equation}
\hat\rho=\exp{[-\beta(\hat H-\mu\hat N-\sum_m \xi_m {\hat
I_m})]}\label{rhohat}.
\end{equation}
As in ordinary equilibrium thermodynamics, the Lagrangian multiplier
associated with the constraint on $N$ is $\mu$, the global chemical potential.
The intensive variables $\xi_m$ are
Lagrangian multipliers associated with the constraints on the currents.
Notice that because of current conservation there are only $M-1$
independent current constraints, and hence one $\xi_m$ can be chosen freely.
It turns out to be convenient
to choose $\xi_d=0$, and we do so henceforth.
Associated with the constraint on $U$ is the Lagrangian multiplier $\beta$.
In equilibrium $\beta$ is the inverse temperature;  we shall shortly present
several arguments why this continues to be true here.

For clarity let us consider the case of non-interacting electrons. These
occupy any of a complete set of single-particle eigenstates
$|\psi_\alpha\rangle$ of $\hat H$ and $\hat I_m$.  The density matrix
$\hat\rho$ above leads to a thermal occupancy of these single-particle
states given by
\begin{equation}
f_{\alpha}={1\over\exp{[\beta(\epsilon_{\alpha}-\mu-\sum_m\xi_m
i_{m,\alpha})]}+1}.\label{occ}
\end{equation}
To make this more transparent, let us look at this in the case
of a two-terminal system, using scattering states $|\psi_{mnk}\rangle$
for the single-particle basis.  In the two-terminal case we can drop the
terminal index $m$, and understand that $k>0$  ($k<0$) corresponds to states
injected by the source (drain).  These can also be called right- and
left-movers, respectively.
In an ideal system with no backscattering
[$t_{n'k',nk}=\delta_{n'n}\delta_{k'k}$],
these states carry currents $i_{nk}$ from end to end, and
the distribution becomes
\begin{equation}
f_{nk}={1\over \exp[{\beta(\epsilon_{nk}-\mu-\xi i_{nk})}]+1},
\label{occ2term}
\end{equation}
with $\xi\equiv\xi_s$.
With only one subband ($n=0$) occupied, this is similar to the LB
result; at zero temperature, states are occupied up
to an energy $\mu+\xi i_{0k}$ (with $k$ here the highest occupied
state on an edge).  This is pictured in Fig.~\ref{MEAfig}(a).
Because the current has opposite signs for positive and negative $k$,
right- and left-moving states are occupied up to different energies.
The combination $\mu+\xi i_{0k}$ acts in this case
like an effective local chemical potential.
In the general case with several subbands occupied, however,
states are occupied up to different
energies in each subband.  [See Fig.~\ref{MEAfig}(b).]
At zero temperature the states in
subband $n$ are occupied up to $\mu+\xi i_{nk}$ which depends on
the current carried by the highest occupied state in this subband.
This current is in general different in different subbands,
and so the distributions $f_{nk}$
cannot be described in terms of local chemical potentials.
Said another way, if one insisted on defining local chemical potentials,
there would need to be a different `local chemical potential' for each
subband.

\subsection{Calculating voltages}

Within the reservoir model it is quite clear that ordinary
voltage measurements at a contact correspond to the local
chemical potential: $V_m = \mu_m/e$.  (See, for example,
\onlinecite{ButtikerRev}.) As we have
emphasized above, the distributions derived from the MEA
cannot in general be described in terms of a local chemical
potential, and clearly the LB prescription for finding
$V_m$ cannot apply.  Nor does the MEA itself give some
procedure for determining voltages.  The approach we take
to calculate voltages comes from the following physical
picture:  a voltmeter determines the voltage differences
between two terminals by measuring the work required to move
a small amount of charge from one to the other.
If moving some charge $\delta Q$ takes a work $\delta W$,
then the voltage difference is the ratio
$\delta W/\delta Q$.  The problem is then
to calculate this work.

The work required to move reversibly between equilibrium states is
given by changes in thermodynamic potentials ({\it e.g.},
the Helmholtz free energy, when the temperature is constant).
In general potentials cannot be defined in nonequilibrium systems,
which ordinarily involve dissipation.  The
problem of steady-state transport in mesoscopic systems,
however, is a special case of nonequilibrium thermodynamics.
In the device there is no inelastic scattering; this, and
the steady-state condition, permit us to define thermodynamic
potentials of the electron distribution \cite{Tykodi}.
For example, the information entropy is equivalent to an
ordinary thermodynamic entropy, as mentioned above and explained
below; and so
here we can define the Helmholtz free energy as usual, $F=U-TS$.
The work $\delta W$ done on the system at constant temperature
is then equal to the change in free energy,
\begin{equation}
\delta{F} = \mu\,\delta\!N + \sum_m\xi_m\,\delta\!I_m.
\label{deltaF}
\end{equation}
That is, at constant temperature the free energy is a function of the electron
number and of the net current at each terminal.  However, the variables
most easily varied in the theory are the intensive variables
${\mu,\xi_s,\xi_1,\dots,\xi_{M-2}}$.  Varying any of these in general
changes the occupancy $f_{\alpha}$ of incoming states at each terminal,
by Eq.~(\ref{occ}).  Here for definiteness we will use the representation
given by the scattering states $|\psi^+_{mnk}\rangle$ labeled by
$\alpha=mnk$, with occupancies $f_{mnk}$.
In what follows it is useful to define a quantity $N_m$ which is the number of
occupied scattering states entering at terminal $m$.
Thus $N_m=\sum_{nk}f_{mnk}$ and the total electron number in the device
is $N=\sum_m N_m$.

Let $\eta$ refer to any of the independent variables
${\mu,\xi_s,\xi_1,\dots,\xi_{M-2}}$.  Let $\delta\!I_m^{\eta}$ denote
the change in the net current entering at terminal $m$ when variable
$\eta$ is varied while the other independent variables are held fixed.
Similarly let $\delta\!N^{\eta}_{m}$ be the corresponding change in $N_m$.
Then changing $\eta\rightarrow\eta+\delta\eta$ produces a free energy change
\begin{equation}
\delta F^{\eta}= \sum_m \left( \mu \delta\!N^{\eta}_m +
               \xi_m \,\delta\!I^{\eta}_m \right) .
\label{deltaF2}
\end{equation}
We obtain the potentials $V_{m}$ at the terminals by
interpreting this free energy change as the work done to add $\delta\!N^\eta_m$
electrons against the voltage $V_m$ at the terminals.  At each terminal the
energy cost is $e\delta\!N_m^{\eta} V_m$.  Thus we can also write
\begin{equation}
\delta F^{\eta} = \sum_m e\, \delta\!N^{\eta}_m V_m.
\label{deltaF3}
\end{equation}
Equating Eqs.~(\ref{deltaF2}) and (\ref{deltaF3}) for each of the $M$
variables $\eta$ gives a set of $M$ linearly independent equations,
\begin{equation}
\sum_{m} \delta\!N^{\eta}_{m} (eV_{m}-\mu) = \sum_m \xi_m \delta\!I^{\eta}_m,
\label{lineq}
\end{equation}
which are to be solved for the unknown terminal voltages $V_{m}$.
The non-local resistance measured between two terminals $m$ and $m'$
is then $R_{sd,mm'}=(V_m-V_{m'})/I$.  The matrix $\delta\!N_m^{\eta}$
on the left-hand side of Eq.~(\ref{lineq}) is invertible, and the resulting
potentials are then automatically given relative to the global chemical
potential $\mu$.

\subsection{Example}

In this section we will illustrate the approach outlined above
by finding the resistance of an ideal two-terminal system, and
then extend the result to the ideal multi-terminal case.
First consider a two-terminal device.
For this we can use eigenstates which satisfy periodic boundary
conditions on a length $L$ along the device.  This particular choice
of boundary condition gives a simple density of states, but is of
no other significance. The states are
labeled by a subband index $n$ and wavevector $k$.
Again in this case we can drop the terminal
subscripts and understand that states with $k>0$ are injected
by the source and $k<0$ by the drain.  We choose $\xi_d=0$ and
write $\xi$ for $\xi_s$.  The net current $I_s$
entering at the source is just the total current $I$ through
the device.  Here we consider only one
simple case, for which the states have energy
$\epsilon_{nk}=\epsilon_n + \hbar^2 k^2/2m^{*}$ and  current
$i_{nk}=e\hbar k/ m^*L$.  This can represent 1D
transport, or a Hall bar with a parabolic transverse
confinement.  (In the latter case,
$\hbar^2/2m^{*}\rightarrow \omega_0\ell^2/2$, where $\omega_0$
is the curvature of the confining potential and $\ell=\sqrt{\hbar c/eB}$
is the magnetic length.)
By Eq.~(\ref{occ2term}), the occupancy $f_{nk}$ depends on energy and
current in the combination $\epsilon_{nk}-\xi i_{nk}$.  This is
\begin{eqnarray}
&&\epsilon_n + {\hbar^2 k^2\over 2m^{*}} - \xi {e\hbar k \over m^{*}L}
\nonumber\\
&=&\epsilon_n + {\hbar^2\over2m^{*}}(k-{e\xi\over\hbar L})^2 -
{e^2\xi^2\over 2m^{*}L^2}.
\label{combo}
\end{eqnarray}
Thus the occupancy can be written
\begin{equation}
f_{nk} = f({\tilde\epsilon}_{nk}-\tilde\mu) \equiv
1/(e^{\beta(\tilde\epsilon_{nk}-\tilde\mu)}+1),
\label{f1}
\end{equation}
where we have defined
\begin{mathletters}
\begin{eqnarray}
\tilde\epsilon_{nk} &=& \epsilon_n + {\hbar^2\over2m^{*}}(k-\tilde\xi)^2,
\label{tildeepsilon}\\
\tilde\xi &=& {e\xi\over\hbar L},\\
\tilde\mu &=& \mu+{\hbar^2{\tilde\xi}^2\over2m^{*}},
\label{tildemu}
\end{eqnarray}
\end{mathletters}
for reasons which will shortly be apparent.
The total number of electrons occupying current-carrying states in
the device is
\begin{equation}
N = \sum_{nk} f(\widetilde\epsilon_{nk}-\widetilde\mu)
  = \sum_n  \int_{\epsilon_n}^{\infty} d{\widetilde\epsilon}\,
\rho_n({\widetilde\epsilon})
f({\widetilde\epsilon}-\tilde{\mu}),
\label{N}
\end{equation}
where in the limit of large $L$ we can take the $k$'s and hence
${\widetilde\epsilon}_{nk}$ to be quasicontinuous, and define
for the latter a 1D density of states,
\begin{equation}
\rho_n(\tilde{\epsilon})=(L/\pi)
[2\hbar^2(\tilde{\epsilon}-\epsilon_n)/m^{*}]^{-1/2}.
\label{dos}
\end{equation}
{}From the integral form in
Eq.~(\ref{N}) we see that $N$ depends on $\mu$ and $\xi$ only
in the combination $\widetilde\mu$ given by Eq.~(\ref{tildemu}).
Thus here $\widetilde\mu$ acts like a current-dependent chemical
potential controlling the number of occupied current-carrying states.
The total current carried by the occupied states is
\begin{equation}
I=\sum_{nk} i_{nk} f(\widetilde\epsilon_{nk}-\widetilde\mu).
\end{equation}
Notice that $f_{nk}$ is symmetric about $k=\widetilde\xi$.  That is,
the two states in subband $n$ at
$k_{\pm} =\widetilde\xi \pm
[2m^{*}(\widetilde\epsilon-\epsilon_n)/\hbar^2]^{1/2}$
have the same value of $\widetilde\epsilon_{nk}=\widetilde\epsilon$ and hence,
from Eq.~(\ref{f1}), the same occupancy.
The current carried by these two states is then
\begin{equation}
i_{n}(k_+) + i_n(k_-) = e\hbar (k_+ + k_-)/m^*L = 2e\hbar\widetilde\xi/m^*L.
\end{equation}
Thus the total current is
\begin{equation}
I = {e\hbar\widetilde\xi\over m^*L} \sum_{nk}
f(\widetilde\epsilon_{nk}-\widetilde\mu)
  = {e\hbar\over m^*L} \widetilde\xi N.
\label{I}
\end{equation}

With these expressions it is now straightforward to compute the
current-voltage relations by inverting Eqs.~(\ref{lineq}).  In
this two-terminal example, it is easier to use $\widetilde\mu$
and $\widetilde\xi$ as independent variables instead of the
original pair $\mu, \xi$.  Then for this example there are
two Eqs.~(\ref{lineq}), labeled by $\eta=\widetilde\mu, \widetilde\xi$:
\begin{equation}
\left(
\begin{array}{cc}
\delta\!N_s^{\tilde{\mu}} & \delta\!N_d^{\tilde{\mu}} \\
\delta\!N_s^{\tilde{\xi}} & \delta\!N_d^{\tilde{\xi}}
\end{array} \right) \left(
\begin{array}{c}
eV_s^{\phantom{\tilde\mu}} - \mu \\
eV_d^{\phantom{\tilde\mu}} - \mu
\end{array} \right)
= \xi
\left(
\begin{array}{c}
\delta\!I^{\tilde{\mu}} \\
\delta\!I^{\tilde{\xi}}
\end{array} \right) .
\label{2term}
\end{equation}
These equations contain $\delta\!N_m^{\eta}$, the change in
the occupation number $N_m$ of states which carry current into terminal
$m$.    We are interested in the changes in these which result
when $\eta=\widetilde\xi, \widetilde\mu$ are changed infinitesimally.
Rather than computing these directly, consider a function
\begin{equation}
Q(\widetilde\mu,\widetilde\xi;k_0) =
\sum_{nk} f(\widetilde\epsilon_{nk}-\widetilde\mu) {\rm sgn}(k-k_0).
\label{Q}
\end{equation}
Notice that $Q$ depends on $\widetilde\xi$ via
$\widetilde\epsilon_{nk}$, which, by Eq.~(\ref{tildeepsilon}),
depends on $\widetilde\xi$.
The quantity ${1\over2}(N\pm Q)$ gives the number of occupied states
with $k\gtlt k_0$. We ultimately seek the changes in number of
occupied states
with $k\gtlt0$ due to infinitesimal changes $\delta\widetilde\mu$ and
$\delta\widetilde\xi$. But the changes in occupation of these states
are the same as the changes in occupation numbers of
states with $k\gtlt k_0$, as long as $k_0$ is not too close to the
edge of occupied $k$'s (that is, as long as states near wavenumber
$k_0$ are either fully occupied or unoccupied).  Thus we can write
\begin{equation}
\delta\!N^{\eta}_{s,d} = {1\over2}(\delta\!N^{\eta} \pm \delta Q^\eta),
\label{deltaNsd}
\end{equation}
where the upper (lower) sign is for $m=s$ ($d$), and $\eta$
represents variations in either $\widetilde\mu$ or $\widetilde\xi$.
Let us choose $k_0=\widetilde\xi$.
[Then, more precisely, Eq.~(\ref{deltaNsd}) is true as long as the region
of $k$-space where $f(\widetilde\epsilon_{nk}-\widetilde\mu)$ is very close to
unity (at low temperatures) brackets both $\widetilde\xi$ and $k=0$.  This will
be discussed further in Part~2, since this point leads
to a breakdown in quantization in point contacts.]
Clearly $Q(\widetilde\mu,\widetilde\xi; k_0=\widetilde\xi)=0$.
We seek
\begin{equation}
\delta Q \equiv Q(\widetilde\mu+\delta\widetilde\mu,
\widetilde\xi+\delta\widetilde\xi; \widetilde\xi)
= \sum_{nk} {1\over e^{\beta({\widetilde\epsilon_{nk}}'-
\widetilde\mu-\delta\widetilde\mu)}+1} {\rm sgn}(k-\widetilde\xi),
\label{Qa}
\end{equation}
where
${\widetilde\epsilon_{nk}}'=\epsilon_n +
\hbar^2(k-\widetilde\xi-\delta\widetilde\xi)^2/2m^{*}$.
Since ${{\widetilde\epsilon}_{nk}}'$ is symmetric about
$\widetilde\xi+\delta\widetilde\xi$, the only contribution
to the sum over $k$ is for
$\widetilde\xi\le k \le \widetilde\xi+2\delta\widetilde\xi$,
so that
\begin{equation}
\delta Q =
\sum_n \sum_{k=\widetilde\xi}^{\widetilde\xi+2\delta\widetilde\xi}
\left[ e^{\beta(\widetilde\epsilon_{nk}'-\widetilde\mu-\delta\widetilde\mu)}
 +1 \right]^{-1}
\approx {L\over\pi} \delta\widetilde\xi \sum_n
\left[e^{\beta(\epsilon_n-\widetilde\mu)}+1\right]^{-1}
\label{deltaQ}
\end{equation}
plus terms of second order in the quantities $\delta\widetilde\xi$,
$\delta\widetilde\mu$.  That is, $\delta Q^{\widetilde\xi}=\delta Q$
and $\delta Q^{\widetilde\mu}=0$.
{}From the expression for $N$ given in Eq.~(\ref{N}) it is evident that
varying $\widetilde\xi$ and $\widetilde\mu$ changes $N$ by
\begin{equation}
\delta\!N = - \delta\widetilde\mu \sum_n \int_{\epsilon_n}^{\infty}
d\widetilde\epsilon \rho_n(\widetilde\epsilon)
{\partial f(\widetilde\epsilon-\widetilde\mu)\over \partial\widetilde\epsilon},
\label{deltaN}
\end{equation}
that is, $\delta\!N^{\widetilde\mu}=\delta\!N$ and
$\delta\!N^{\widetilde\xi}=0$.
Combining Eqs.~(\ref{deltaN}), (\ref{deltaQ}), and (\ref{deltaNsd}),
we find
\begin{eqnarray}
\delta\!N_s^{\widetilde\mu} &=&  \delta\!N_d^{\widetilde\mu}
= {1\over2}\delta\!N,\nonumber\\
\delta\!N_s^{\widetilde\xi} &=& - \delta\!N_d^{\widetilde\xi}
= {1\over2}\delta Q.
\label{finaldN}
\end{eqnarray}
{}From Eq.~(\ref{I}) we obtain
\begin{eqnarray}
\delta\!I^{\widetilde\mu} = {e\hbar\over m^* L}\widetilde\xi
\delta\!N,\nonumber\\
\delta\!I^{\widetilde\xi} = {e\hbar\over m^* L}N \delta\widetilde\xi.
\label{finaldI}
\end{eqnarray}

It is now a simple matter to invert the matrix in Eq.~(\ref{2term})
and solve for $V_{s,d}$.  We write the final answer as the
two-terminal conductance
\begin{equation}
G=I/(V_s-V_d)= {e^2\over h} \sum_n
{1\over e^{\beta(\epsilon_n-\widetilde\mu)}+1}.
\label{conductance}
\end{equation}
If $\widetilde\mu$ exceeds the band minima of the first $j$
subbands (or Landau levels), then at low temperatures
$G=je^2/h$.  Finite-temperature corrections are exponentially
small.  This exact result is true for a system without backscattering
regardless of the size of the voltage or current.
This occurs because at large currents it is quite possible for states in
one subband to be occupied up to energies above those of empty states in
other subbands---not because relaxation has failed to occur, but because
according to the MEA this is the steady-state (although highly
nonequilibrium) result.
Exceptions to quantization occur when either (a) the
current grows so large that a subband is occupied only for
carriers moving in one direction (this is connected to the
saturation in quantum point contacts \cite{van Houten}, and will be
discussed further in part~2); or (b) the
current is large enough to induce breakdown in the sample as a
consequence of other dissipative mechanisms \cite{Cage}.
For ordinary samples exhibiting the IQHE, neither of these
occurs, and the approach described above provides an explanation
of the extremely accurate quantization observed outside the
linear response regime.
This ability to explain the high-precision IQHE experiments is
non-trivial, and lends credence to the postulate of the MEA.

The above example was explained in detail to show
how our approach is applied in general.  In this particular
case of a two-terminal system, a simplification is possible.
Since $N$ depends on $\widetilde\mu$ and not $\widetilde\xi$,
changing the latter while holding the former constant corresponds
to moving a charge $e\delta Q/2$ from one
terminal to the other (or from edge to edge in a quantum Hall
sample).  The work required to do this (at constant $N$) is
$\delta F^{\widetilde\xi}=\xi \delta\!I^{\widetilde\xi}$.
Hence the voltage difference is
\begin{equation}
e(V_s-V_d) = {\delta F^{\widetilde\xi}\over e\delta Q/2},
\label{simpler}
\end{equation}
which gives the same result.

It is easy to extend the above calculation to an ideal multi-terminal
IQHE system. (The general formalism is valid regardless
of the presence of a magnetic field.)
Suppose there are four identical terminals, as shown in
Fig.~\ref{devicefig}.  Then all of the current leaving the source flows
along the lower edge to terminal~1.  For the net current through
terminal~1 to vanish, it is necessary that $\xi_1=\xi_s$.  Similarly
$\xi_2=\xi_d=0$.  The work required to transfer a certain charge from
terminal 1 to 2 is then precisely that required to move the same
charge from $s$ to $d$ in the two-terminal calculation.  Hence the
two-terminal conductance calculated above becomes here the Hall
conductance.

Finally, we mention that the parabolic example is special only
in that it can be solved analytically.
We have numerically studied non-parabolic energies $\epsilon_{nk}$ and
multi-terminal systems
and find in these cases that the accuracy of the quantization is limited
only by the numerical accuracy.  This work will be reported in
part 2.

\section{Discussion}

\subsection{Density matrix}
\label{distributions}

The density matrix and distribution which result from the maximum entropy
approach, and which lie at the heart of our ability to explain the
high-precision IQHE experiments, are unusual, but not unheard of in
the literature.  The distribution $f_{\alpha}$ in Eq.~(\ref{occ})
was proposed earlier by Heinonen and Taylor \cite{HeinonenTaylor},
and more recently by Ng \cite{Ng}.  These authors argued
argued that in a device without dissipation it should be
possible to define a free energy which, presumably, would be
minimized.  The process of minimization, subject to the current
constraint, is formally identical to the MEA's maximization of
information entropy, and leads to the same distribution.  In
this paper
we have obtained this result on considerably more fundamental
grounds---assuming only 
that the postulates of the maximum entropy approach to nonequilibrium
statistical mechanics are, in fact, correct.

Here we will seek some understanding of the density matrix
Eq.~(\ref{rhohat}) and distribution Eq.~(\ref{occ}) from other
viewpoints.
Note first of all that this density matrix has the general form which
Hershfield recently showed should exist on quite general grounds in
steady-state nonequilibrium systems \cite{Hershfield}.  In his
rather more conventional approach to nonequilibrium statistical
mechanics, it is quite evident that $\beta$ in Eq.~(\ref{occ})
is indeed $1/k_BT$, where $T$ is the temperature.  This was not
obvious in the MEA, but based on Hershfield's work we can make
the same identification, in the case of steady-state mesoscopic
transport.  It also then follows that the information and
thermodynamic entropies are equivalent [with $c$ in Eq.~(\ref{entropy})
set to $k_B$].

The distributions Eq.~(\ref{occ}) are obviously quite different
from the LB distributions.  The latter follow in a very straightforward
way from the model of terminals as reservoirs, and
are widely considered valid by their ability to describe non-local
resistances in many low-current experiments, as well as by the
appealing simplicity and clarity of the reservoir model itself.
How can this difference in distributions be understood physically?
The LB distributions can be derived from linear response theory, with a
certain assumption about how the system is driven.
One can apply something very similar to the reservoir idea
in a very precise calculation by modeling the leads as ideal and
supposing that far from the device
there is in each lead a well-defined electrochemical potential.
A.D. Stone and coworkers have shown that this assumption leads
{\em in linear response} to the multi-terminal LB
formalism \cite{ADStone,BarangerStone}.
We should note, however, that this work has been criticized by
Landauer, based on their use of leads of constant cross section,
which do not have the geometrical spreading he believes is
necessary for the reservoir picture \cite{LandauerGW}.
Even so, while indeed no large reservoir is invoked in the
work of Stone {\it et al.}, a very
similar idea---that far from the device a probe can be described
as a system at constant potential---is at the heart of this
calculation, entering as a boundary condition.
(We should also note that since in principle linear response
can be calculated using equilibrium distributions, the success
of the linear theory in predicting low-current properties does
not imply that the linear distributions are correct.)

We emphasize that built into Stone {\it et al.}'s and related calculations is
the model of terminals as entities described
by local chemical potentials.  This model fits neatly into the
most common way to approach nonequilibrium problems:  assuming
that there are two large reservoirs, each in equilibrium, and
that transport (say) between them occurs when they are connected
by a small channel.
This is a very familiar approach to nonequilibrium
statistical mechanics.
And yet it amounts to an assumption as to how the system is
driven.  In typical transport experiments in the IQHE, say,
a constant current source is connected to the sample.  Using
the reservoir picture amounts to a model of how a current
source (when connected to a mesoscopic device) actually
drives the current.  Perhaps it is valid in linear response
to model the current source as two leads at different potentials.
Even if valid in linear response, it is not {\it a priori} clear
that the picture should be valid at large currents.  In fact, the
failure of the LB approach at large currents suggests not.

Let us examine this in more detail.  Suppose that the current
source itself can be viewed as an object
which gives rise to different potentials in the physical leads.
Then the current is carried down long macroscopic lengths of lead
until it is injected into the device through the terminals.  Perhaps
along the macroscopic wire the local potential
changes gradually and smoothly (due to elastic and inelastic scattering);
that is, perhaps the distribution
locally can be described by a local chemical potential.  Does
this picture work right up to the vicinity of the mesoscopic
device?  In fact, there are several reasons to think not.

The simplest way to approach this is in the approximation
of the Boltzmann approach to transport \cite{Mahan}.  Consider an
ordinary dissipative conductor carrying a low current.
Let us suppose
that the current is driven by reservoirs held at two
different potentials.  Then at the reservoirs the
Boltzmann distribution $f({\bf r},{\bf k},t)$ will have the LB
form.  But a perfectly standard Boltzmann calculation
shows that far from the reservoirs
the distribution evolves into a current-dependent
form identical (in first order) to that derived
from the MEA [Eq.~(\ref{occ})], and different from
the first-order LB distributions.
Suppose that the ends of an effectively
one-dimensional conductor (at $x=\pm L/2$) are held
at different electrochemical potentials $\mu\pm\Delta\mu/2$.
The distribution is then labeled by a wavenumber $k$ plus
a subband index $n$:  $f=f(x,n,k,t)$.
Suppose further that this gives rise to a uniform electric
field $E=\Delta\mu/eL$ through the conductor.  Then in
steady-state and one dimension the Boltzmann equation becomes \cite{Mahan}
\begin{equation}
v{\partial f\over\partial x} + {\partial k\over\partial t}
{\partial f\over\partial k} = \left({df\over dt}\right)_{\rm coll}.
\label{boltzmann}
\end{equation}
We want the lowest-order (in $\Delta\mu$) solution to this,
in the relaxation time approximation \cite{Mahan}.
Since the conductor's ends are held at definite local chemical
potentials, the distribution takes the LB form at the ends:
\begin{equation}
f(\mp L/2,n,k,t)=1/(\exp[\beta(\epsilon_{nk}-(\mu\pm\Delta\mu/2))]+1).
\label{bc}
\end{equation}
The upper sign is for positive $k$ and the lower for negative
$k$.  Eq.~(\ref{boltzmann}) is a completely standard Boltzmann problem, with
the only wrinkle provided by the boundary conditions, Eq.~(\ref{bc}).
To linear order this problem is solved by
\begin{equation}
f = f_0 + \Delta\mu {\partial f_0\over\partial\epsilon}
\left[ \left( {\hbar k \tau \over mL}\mp {1\over2}\right)
e^{-m(x\pm L/2)/\hbar k \tau} - {\hbar k \tau \over m L}
\right]
\label{f2}
\end{equation}
where again the upper (lower) sign is for $k>0$ ($k<0$).
Here $f_0(\epsilon_{nk})=1/(e^{\beta(\epsilon_{nk}-\mu)}+1)$ is the
equilibrium distribution and $\tau$ is the relaxation
time (which can depend on $k$).
Let us suppose that the relaxation time $\tau$ is
much less than the transit time across the system:
$\tau \ll L/v$, where the speed $v=\hbar k/m$.
Then at distances much farther
than $v\tau$ from the ends the distribution becomes
\begin{equation}
f = f_0 - {\tau \Delta\mu\over e}i_{nk}
{\partial f_0\over\partial\epsilon}.
\label{boltzmanndistr}
\end{equation}
Here $i_{nk}=ev/L$ is the current carried by the state
in subband $n$ with wavevector $k$.
Eq.~(\ref{boltzmanndistr}) is the lowest-order term
in an expansion in current of
\begin{equation}
f_0(\epsilon_{nk}-\xi i_{nk}) =
{1\over \exp[\beta(\epsilon_{nk} -\mu - \xi i_{nk})]+1},
\label{low-order}
\end{equation}
where we have here identified $\xi=\tau\Delta\mu/e$.
Note that this is not equal to the corresponding first-order
LB distribution $f=f_0 \mp (\Delta\mu/2)\partial f_0/\partial\epsilon$.

That is, in a completely typical dissipative conductor,
the Boltzmann distribution is precisely the lowest-order
approximation to the distribution [Eq.~(\ref{occ2term})]
obtained from the maximum entropy calculation.  This
is the case even though we chose the boundaries to
model reservoirs;  the distribution evolves from the
LB form near the ends to the current-dependent form
away from the ends.  Our point with this example
is the following.  Even though the LB distribution
has the authority of widespread usage, perhaps one should
instead typically expect to find current-dependent
distributions carrying steady-state currents into a
mesoscopic system; for the dissipative wires
carrying the current to the mesoscopic device should themselves
typically have such distributions. Note that the Boltzmann equation
gives current-dependent distributions also in the familiar case of a
three-dimensional conductor in the relaxation-time approximation.
In this case the resulting distribution is obtained by displacing
the Fermi surface in the direction of the current, and therefore
here too the occupancy of single-particle states depends on their
current as well as their energy.

There are special cases in which current-dependent distributions
can be obtained in other ways.  We have argued
elsewhere that this should arise by Galilean
transformation in a translationally invariant
dissipationless system \cite{HJ1}.  We also
have given elsewhere a detailed example in which a
mechanism to switch on the current is provided,
and which results in this distribution \cite{HJ3}.
These, and the general result of Hershfield
mentioned earlier \cite{Hershfield}, all support
our identification of $\beta$ as $1/k_BT$,
as well as the notion that these distributions
should be generally expected in steady-state
transport.

If the reservoir picture is only adequate
at low currents, how can one picture the
way a current source drives a large current through
a mesoscopic system?
Perhaps one can think of the current source as forcing
current 
through a region full of scatterers, like someone being
forced to run a gauntlet.  The current source pushes
electrons in;  they scatter into other states.  In
steady state it is reasonable that the occupancies
of the various states will be influenced by their ability to carry current.

Finally, we have emphasized that the distributions resulting from the MEA
(given the constraints of particle number, internal energy, and current)
cannot be described in terms of a local chemical potential.
Nonetheless it is evident from
our calculation of the terminal voltages $V_m$ that the quantity $eV_m$
plays the role of a kind of a local chemical potential.  That is, $eV_m$
is equal to the energy cost required to add an extra particle to the
terminal (more precisely, to occupy an extra incoming state at terminal $m$).
This is clearly not a local chemical potential in the LB sense---the MEA
distributions in terminal $m$ are not of the Fermi-Dirac form with a
local chemical potential $eV_m$.

\subsection{Maximum entropy approach}

Perhaps the most unorthodox part of our calculation
is its use of the maximum entropy approach.  The essential feature
of the MEA (besides the obvious fact of the entropy maximization)
is that observables enter the formalism as constraints.  For example,
here we have treated the current source---the object driving the system
out of equilibrium---merely as something which imposes a constraint
on the total current $I$.  That is, the result of the driving
enters the formalism.  In a typical linear response calculation,
the driver enters as a term in the Hamiltonian (say, a small electric
field).  A difficulty of the MEA
is that it provides no prescription
for how one should determine the constraints.
It seems that one must be guided by the physical
picture.  This has been called ``the basic problem'' with
the maximum entropy approach \cite{LandauerPhysica}.
In many cases it simply is not possible to know what the constraints
are---for example,
in the case of hot electrons in semiconductors it appears that one must
somehow incorporate information about phonon interactions as
a constraint \cite{LandauerPhysica}; and nobody knows how to do this.
But the fact that this approach may be difficult in some problems does
not of course mean that it is always difficult.  In fact, steady-state
mesoscopic transport seems to be ideally suited for this approach.
Guided by the physical picture, we have made the simplest possible
supposition about constraints, and it appears to work.  In particular, since
the current source
is designed to hold the current constant, we simply treat the current
as a constraint.

A second difficulty with the MEA is that often it is difficult to calculate
the microstates are which enter the formalism.
In the case of steady-state mesoscopic transport, this is not a problem.
Since the thermodynamic system is the
mesoscopic device plus ideal leads (in the scattering geometry), it
is quite straightforward to calculate the entire set of microstates.
Here we have emphasized how to do this for noninteracting electrons,
but it is also possible in the interacting case.  We will discuss
this more in Part~2.

It is interesting to note that the LB distribution can also be
obtained from the MEA by a different choice of constraint.
This happens if one assumes that the current source somehow
constrains the particle numbers $N_m$ entering at each terminal,
rather than determining the net currents at each terminal \cite{Ng}.
In the MEA these constraints lead to
Lagrangian multipliers $\mu_m$, and the resulting occupancies are the
LB distributions $f_{mnk}^{LB}$.
Thus one might be tempted to ascribe
the difference between the LB distributions
and those obtained by us to the way
in which the current source is modeled.
At low currents the use of
local chemical potentials can be
justified using linear response theory, viewing the potential difference
as driving the current. This
cannot be extended to high currents.
Since the LB distribution is associated with
an ordinary electrochemical potential at
each reservoir, one might suppose that it
models a voltage source instead of a current
source.
If so, then the $I$-$V$ curve at large currents and voltages
would depend on whether voltage
or (as is usual) current is applied \cite{Ng}.
(In the linear regime, both approaches give the same result.)
In fact, based on our arguments in section~\ref{distributions}
this appears unlikely.  Even if a voltage source is applied
to the ends of the macroscopic wires leading to a device, it
appears that by the time one moves far from the source ({\it i.e.},
gets near the device)  one should expect the distribution to have
evolved to a current-dependent form.

We gratefully acknowledge discussions with C.T. Van Degrift, E. Palm, S.
Girvin, S. Hershfield,
M.~B\"uttiker and A.D.~Stone.  This work was supported in part by the
UCF Division of Sponsored Research, and
by the National Science Foundation under grant DMR-9301433.

\begin{figure}
\caption{Typical energy spectrum of a one-dimensional
mesoscopic device, or a laterally confined two-terminal IQHE system,
and zero-temperature occupancies of the single-particle states
in the Landauer-B\~uttiker
picture.  In this picture the occupancies of the states
entering the device from the source and drain are described by
local chemical potentials $\mu_s$ and $\mu_d$, respectively.
Occupied states are marked by a heavy line.
In (a) $\mu_{s,d}$ exceed the minimum of the lowest $j$ subbands
(or Landau levels), but lie below the minimum of the $(j+1)^{\rm st}$ subband.
In (b) $\mu_s$ (but not $\mu_d$) lies above the minimum of the $(j+1)^{\rm st}$
subband.}
\label{LBfig}
\end{figure}
\begin{figure}
\caption{Schematic representation of a typical four-terminal IQHE device.
Current runs along
the lower edge from $s$ to terminal $1$ and then to $d$.   Similar
currents run along the upper edge.  If the device is ideal, and
all terminals are identical, then in the Landauer-B\"uttiker picture
$\mu_1=\mu_s$ and $\mu_2=\mu_d$ to get zero
net current in terminals 1 and 2.  Similarly in the maximum
entropy approach $\xi_1=\xi_s$ and $\xi_2=\xi_d$.}
\label{devicefig}
\end{figure}
\begin{figure}
\caption{The typical bending of energy levels in an IQHE sample
that occur due to electrostatics, here a combination of a
transverse Hall electric field and confinement at the edges.
(a) Occupied states are only in the first subband, and lie below the
minimum of the next subband.
(b) As pictured, states in only the lowest subband are occupied, even
though some occupied states are higher in energy than empty states
in the next subband.
}
\label{bent}
\end{figure}
\begin{figure}
\caption{Current-dependent occupancies of states in an ideal
two-terminal device, from the maximum entropy method.
Occupied states (at zero temperature) are indicated by
a heavy line.  (a) When only
one subband is occupied, states are occupied up to energies
$\mu+\xi i_{0k}$.  These have different values for the outermost
occupied states with for $k>0$ and $k<0$ (shown at $k$ and $k'$,
respectively).
(b) When more than one subband is occupied, states in different
subbands are occupied up to different energies.  Compare this
with Fig.~1.}
\label{MEAfig}
\end{figure}

\begin{references}
\bibitem{currentoperator} $\hat I_m$ can be defined in terms of an integral
over matrix elements of the current density operator at terminal $m$, averaged
over the lead (see Ref. ~\onlinecite{Ng}).
The current operator so defined commutes with $\hat H$.
\end{references}
\end{document}